\documentclass[groupedaddress,prd,twocolumn,showpacs,floatfix,amsmath,amssymb,aps]{revtex4-2} 
\usepackage[hidelinks]{hyperref}
\usepackage{graphicx}
\usepackage{tikz}
\usepackage{physics}
\usepackage{mathtools}
\usepackage[capitalise]{cleveref}
\usepackage{dsfont}
\usepackage{microtype}
\usepackage{enumitem}
\usepackage{subfigure}

\usepackage[normalem]{ulem}

\newcommand{\FH}[1]{\textcolor{black}{#1}}

\numberwithin{equation}{section}

\begin{document}
\title{Beyond the ensemble paradigm in low-dimensional quantum gravity: \\ Schwarzian density, quantum chaos and wormhole contributions}
\newcommand{\RegensburgUniversity}{Institut f\"ur Theoretische Physik, 
Universit\"at Regensburg, D-93040 Regensburg, Germany}

\author{Fabian Haneder}
\affiliation{\RegensburgUniversity}
\author{Juan Diego Urbina}
\affiliation{\RegensburgUniversity}
\author{Camilo Moreno}
\affiliation{\RegensburgUniversity}
\author{Torsten Weber}
\affiliation{\RegensburgUniversity}
\author{Klaus Richter}
\affiliation{\RegensburgUniversity}
\begin{abstract}

Based on periodic orbit theory, we address the individual-system versus ensemble interpretation of quantum gravity from a quantum chaos perspective. To this end, we show that the spectrum of geodesic motion on high-dimensional hyperbolic manifolds, described by the Selberg trace formula, displays a Schwarzian ($\sinh 2\pi\sqrt{E}$) mean level density. Due to its chaotic classical limit, this quantum system also shows all universal signatures of quantum chaos. These two properties imply a possible duality to Jackiw-Teitelboim-type quantum gravity at the level of a \textit{single system} instead of an \textit{ensemble of systems} like matrix- and SYK models. Moving beyond the universal regime, we show how the full wormhole geometry on the gravity side emerges from the discreteness of the set of periodic orbits. Thereby, we take initial steps towards a duality between gravitational and mesoscopic chaotic quantum systems through the topological, respectively, periodic orbit expansions of their correlators.
\end{abstract}
%
\maketitle

\section{Introduction}
Low-dimensional gravity has been a fruitful area for studying holography in recent years, with much interest arising from the paradigm that quantum gravitational degrees of freedom display characteristic signatures of quantum chaos
in the presence of a horizon
\cite{Cotler2016a}. A prime example of this is Jackiw-Teitelboim (JT) gravity \cite{Teitelboim1983,Jackiw1985}, whose quantization has been shown to be exactly dual to a matrix ensemble with a universal limit, given by Random Matrix Theory (RMT) \cite{Saad2019}. Originally devised to consider specific signatures of chaos for the unitary case of broken time-reversal symmetry (see \cite{Mertens2023} for a recent review), further work clarified how to extend the duality to other universality classes \cite{Stanford2019, yan2022crosscapcontributionlatetimetwopoint,saad_convergent_2022, stanford_mirzakhani_2023, weber_unorientable_2024} and even used it to explore properties of hyperbolic manifolds \cite{Saad2019,blommaert_integrable_2023,Weber_2023}.

JT gravity displays a characteristic density of states \cite{Yang2018} $\sim \sinh (2\pi\sqrt{E})$, a hallmark of systems with a black hole dual \cite{hawking_black_1976}. It arises from the Schwarzian action describing the leading-order (disk) contribution to the 1-point spectral function. Given the expectation of gravitational systems being maximally chaotic \cite{Maldacena2016, Blake2021}, a possible duality requires the presence of general universal signatures of late-time quantum chaos, but a complete duality \FH{implies more; namely, a precise identification of the respective Hilbert spaces and the spectra of the Hamiltonians of the two theories}. Summarizing the state of the art, the celebrated matrix ensemble of \cite{Saad2019} as dual to JT quantum gravity and the approach of \cite{Post2022,Altland2022} that derives the JT correlators from expectation values within a larger ``universe'' theory both required the Schwarzian density as an external input, while the SYK \cite{Maldacena2016a,Jensen2016,Kitaev2018} and related models \cite{Jia2020,Cotler2016a,Garcia2017} display the Schwarzian action, but only match the gravity side in the universal regime \cite{okuyama2024babyuniverseoperatorsdoublescaled}. 

The study of JT gravity has allowed to address problems in more general and realistic theories of gravity \cite{Johnson2021c,cotler_2023_isometric,Shen2024}, among them the inclusion of higher-than-disk topologies in the gravitational path integral, but at the same time it raised the question of how to understand such topologies from the holographic perspective. In particular, wormhole geometries correspond to non-factorizing correlation functions in the dual theory. In JT gravity, the emergence of non-vanishing connected 2-point correlators can be explained by an ensemble interpretation of the dual theory \cite{Marolf2020,Blommaert2020,Blommaert2021,Altland_2021jd}, but in higher dimensions this interpretation is not available and the so-called factorization problem remains~\cite{Maldacena2004}. 

Although the ensemble interpretation is supported by the existence of matrix and SYK-related dual theories, it apparently contradicts the holographic paradigm embodied by the AdS/CFT correspondence, where the duality relates two \textit{individual} systems, see e.g. \cite{Saad2021b,Marolf_2021,marolf2024natureensemblesgravitationalpath}. The study of JT gravity from the individual-system perspective is therefore an active field, with ideas being pursued in three-dimensional gravity \cite{di_ubaldo_ads3rmt2_2023}, constrained matrix models \cite{Blommaert2021}, non-perturbative corrections \cite{Boruch2024} and the effective description as a ``universe'' field theory \cite{Post2022,Altland2022}.

Here, we appeal to the success of periodic orbit theory \cite{gutzwiller1991chaos,haake2001quantum} in explaining both universal \cite{Berry1985a,Sieber2003, Muller2004} and non-universal \cite{Tanner2000,Ho2019} features of quantum chaotic systems, see \cite{richter_semiclassical_2022} for a recent review. Using this semiclassical approach, we explicitly construct an \textit{individual} chaotic quantum system, i.e., high-dimensional hyperbolic dynamics \cite{gutzwiller_geometry_1985}, that possesses the same (Schwarzian) spectral density as JT gravity. As for any quantum chaotic model, the standard semiclassical smoothing defined by averaging over small parameter windows \cite{Muller2004,Altland_2021jd} reproduces the \textit{universal} (RMT) limit of the leading topology term in the JT 2-point correlation function \cite{Cotler2016a,Saad2019}, the so-called ramp contribution, in systems with both broken and preserved time-reversal invariance \cite{Berry1985a}. After establishing the duality at this universal level, we use the discreteness of the set of periodic orbits to recover the full, non-universal gravitational result as given by the double-trumpet topology \cite{Saad2019}. We emphasize that the double-trumpet possesses a non-universal parameter regime that standard semiclassics cannot capture, whereas our approach is equivalent to the JT gravity result at leading topology in any regime. This equivalence appears when a statistical description of the set of periodic orbits is induced by a coarse graining of the classical phase space at a finite action scale.

\FH{As a final note, let us be clear that all the statements made in this work operate at the level of correlation functions of the theories under study only. While we believe that these results are interesting in their own right, and constitute progress on the way to a computationally useful single system dual to JT gravity, we do not claim to have established a full duality between the theories in the more rigorous sense of a one-to-one correspondence between Hilbert spaces.}

\FH{This paper is organized as follows: in \cref{sec:JT}, we briefly introduce JT gravity and review some features that are pertinent to the discussion in the rest of the paper. In \cref{sec:model}, we construct the quantum chaotic model we will be using in \cref{sec:semiclass} to produce the JT gravity one- and two-point functions at leading topology, but only in the case of conjugate-time arguments for the two-point function, i.e. the regime in which standard semiclassics and RMT agree. In \cref{sec:avg}, we will go beyond this result and show how a phase space coarse graining can produce the full JT gravity double trumpet result without requiring the time arguments to be conjugate to each other. Finally, in \cref{sec:genus}, we give an outlook on how the coarse graining introduced in \cref{sec:avg} could give rise to a genus-like expansion of correlation functions, as expected in JT gravity.}

\section{Jackiw-Teitelboim gravity}\label{sec:JT}
We begin \FH{by} introducing JT gravity as a two-dimensional theory of dilaton gravity defined by the action
\begin{equation}
  \begin{aligned}
  S_{JT}&[g,\phi]=-\frac{S_0}{2\pi}\left[\frac{1}{2}\int_\mathcal{M}\sqrt{g}R+\int_{\partial\mathcal{M}}\sqrt{h}K\right]\\
  &-\left[\frac{1}{2}\int_\mathcal{M}\sqrt{g}\phi(R+2)+\int_{\partial\mathcal{M}}\sqrt{h}\phi(K-1)\right],
  \end{aligned}\label{eq:S_JT}
\end{equation}
where $S_0$ is a large parameter setting the entropy scale and suppressing topology change, $\mathcal{M}$ is a two-dimensional Riemannian manifold, $\partial\mathcal{M}$ its boundary, $g$ and $R$ the metric determinant and Ricci scalar on $\mathcal{M}$, $h$ and $K$ the induced metric and curvature on $\partial\mathcal{M}$ and $\phi$ a complex scalar field called the dilaton. Equation~(\ref{eq:S_JT}) arises universally as the near-horizon geometry of higher-dimensional near-extremal black holes, see e.g. \cite{Verheijden_2021}.

In its quantized form, the theory is defined through the $n$-boundary connected partition functions 
\begin{equation}
  \ev{Z(\beta_1)\ldots Z(\beta_n)}^{(\text{c})}\simeq\sum_{g=0}^\infty e^{(2-2g-n)S_0}Z_{g,n}(\beta_1,\ldots,\beta_n), \label{eq:correlators}
\end{equation}
expressed as a topological (genus) expansion. The right-hand side in \cref{eq:correlators} is computed by path integrating over the action \eqref{eq:S_JT}, whose topological term (the first bracket) induces the genus decomposition through the Gauß-Bonnet theorem. \FH{Depending on whether we allow unorientable manifolds or not, the sum ranges solely over integer (without unorientable manifolds) or additionally over half-integer (with unorientable manifolds) genus.} The left-hand side sets the boundary conditions, namely that one should integrate over connected manifolds with $n$ asymptotically AdS boundaries of renormalized lengths $\beta_1,\ldots,\beta_n$ (see \cite{Goel2020} for an overview). One can compute the functions $Z_{g,n}$ exactly for all orders $n$ and genus $g$ \cite{Saad2019} \FH{in the orientable case}, but in this \FH{paper}, we will \FH{mostly} focus on the disk and double trumpet geometries, $g=0$ and $n=1,2$, respectively. These two correlators depend directly on integrating the boundary reparameterization mode, $w(u)$, against the emergent Schwarzian action \cite{maldacena2016b}, 
\begin{equation}
  S[w,\phi]=-\frac{S_0}{2\pi}\int du \phi_r(u)\text{Sch}(w,u) \, .
\end{equation}
It is the simplest action invariant under $SL(2,\mathds{R})$, also describing the low energy regime of the SYK model \cite{Maldacena2016a,Kitaev2018a}. Integrating out the Schwarzian action leads, firstly, to the well-known disk partition function
\begin{equation}
  Z_{0,1}(\beta)=Z^{\text{d(isk)}}_{\text{Sch}}(\beta)=\int_0^\infty dE e^{-\beta E}\rho_0(E),\label{eq:disk}
\end{equation}
depending on the characteristic density of states 
\begin{equation}
  \rho_0(E)=\frac{1}{4\pi^2} \, \sinh(2\pi\sqrt{E}) \, .
  \label{eq:sinh}
\end{equation}
Secondly, it yields the wormhole (double-trumpet) contribution to the 2-point function,
\begin{equation}
\label{eq:Z02}
Z_{0,2}(\beta_1,\beta_2)=\int_0^\infty b\,db~Z^{t}_{\text{Sch}}(\beta_1,b)Z^{t}_{\text{Sch}}(\beta_2,b)
\end{equation}
in terms of the characteristic trumpet partition function
\begin{equation}
Z^{\text{t(rumpet)}}_{\text{Sch}}(\beta,b)=\frac{1}{\sqrt{\pi}\beta^{1/2}}e^{-b^2/\beta}.\label{eq:trumpet}
\end{equation}

\FH{The integral in Eq. \eqref{eq:Z02} can be performed exactly. In the case $\beta_1=\beta-it, \beta_2=\beta+it$, \eqref{eq:Z02} computes the spectral form factor and, to leading order in $\beta/t$, this yields the well-known ramp, a telltale sign of chaotic dynamics,
\begin{equation}\label{eq:ramp}
    Z_{0,2}(\beta-it,\beta+it)=\frac{t}{4\pi \beta}+\order{\frac{\beta}{t}}.
\end{equation}
As we will see in \cref{sec:semiclass}, standard semiclassics is able to capture \eqref{eq:ramp}, but not \eqref{eq:Z02}.}

\section{Particle on a high-dimensional hyperbolic manifold}\label{sec:model}
\begin{figure}[ttt]
\centering
\includegraphics[width=0.86\linewidth]{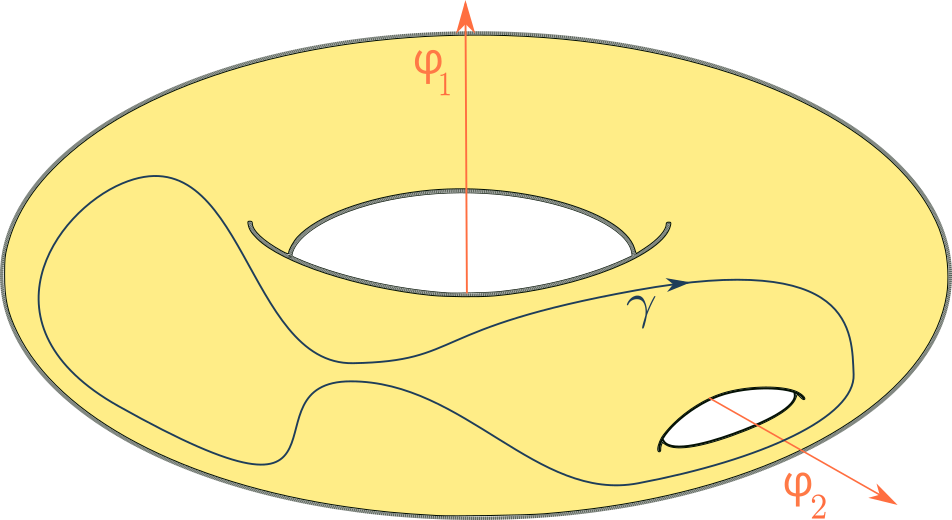}
\caption{\FH{An example of the class of systems we consider in $f=2$ dimensions. If the manifold is embedded in a higher-dimensional space, Aharonov-Bohm fluxes $\varphi_1,\varphi_2$ piercing the handles of the manifold can be added, causing periodic orbits $\gamma$ to pick up phases $\chi_\gamma$ related to the winding number $n_{1,2}(\gamma)$ around the fluxes, thereby breaking time-reversal invariance in the system.}}
  \label{fig1}
\end{figure}

\FH{In this section, we show that} these central quantities, $Z_{0,1}$ and $Z_{0,2}$, can be exactly reproduced by an individual quantum chaotic, non-gravitational, dynamical system. We consider a free particle of mass $M=1/2$ moving on an $f$-dimensional Riemannian manifold $\mathcal{K}$ of constant negative curvature $R=-2$ (or correspondingly, with curvature radius $L$ set to unity). This class of systems is fully chaotic \FH{if $f\geq3$, or if $f=2$ and $\mathcal{K}$ has at least genus 2. In the latter case, the system is known as the Hadamard-Gutzwiller model, see e.g. \cite{aurich_periodic_1988}, and \cref{fig1} for an illustration of an example of the kind of system we consider.} Quantization relies on the Gutzwiller trace formula \cite{gutzwiller1991chaos}, which provides an approximation for e.g. the \emph{quantum} partition function using properties of the system's \emph{classical} periodic orbits. Such systems can be viewed as prototypical examples of chaotic systems, being among the few systems that can be rigorously proven to be classically chaotic. While our presentation in the following refers to a class of systems, all our claims are valid at the level of a single representative system, meaning a concrete manifold ${\cal K}$ of a specific dimension $f$. In the case at hand, Gutzwiller \cite{gutzwiller_geometry_1985} found that the semiclassical approximation is identical \cite{Balazs1986} to the Selberg trace formula (STF) relating the spectrum of the Laplacian $\Delta$ on $\mathcal{K}$ to the length spectrum of the set $\Gamma$ of prime geodesics $\gamma$ on $\mathcal{K}$. 

Using dimensionless variables, the STF is given by \cite{randol_selberg_1984}
\begin{equation}
    \sum_{n=0}^{\infty} h(p_n)=\, \mathcal{V}\int_0^\infty\hspace{-0.7em} h(p)\Phi_f(p)dp + \sum_{\gamma\in\Gamma}\sum_{m=1}^{\infty}\frac{\chi_{\gamma}^{m}l_\gamma g(ml_\gamma)}{S_f(m,l_\gamma)}, 
\label{eq:STF}
\end{equation}
where we parameterized the eigenvalues of $\Delta$ as $\lambda_n=(p_n^2\!+\!(f\!-\!1)/4)$. \FH{$h(p)$ is an even function on the spectrum of $\Delta$ that} satisfies certain \FH{standard} technical conditions on its large-$p$ behavior and analytic structure, \FH{which will be met by our choice below, but which we will not report for clarity of the presentation}. $\mathcal{V}$ is the volume of $\mathcal{K}$ and $\Phi_f(p)$ is the \emph{Plancherel measure} of $\mathcal{K}$, a quantity originating in the group theoretical underpinnings of the STF. In the present context, importantly, it plays the role of the mean spectral density or Weyl term. The first sum ranges over prime geodesics of $\mathcal{K}$ (i.e.\ the periodic orbits of the dynamical system with $\hat{H}=-\Delta$), the second accounts for their repetitions. Lastly, $l_\gamma$ is the length of the $\gamma$-th orbit (in units of $L$), $l_\gamma/S_f$ its stability amplitude, and $g(l)$ is the Fourier transform of $h(p)$. $\chi_\gamma$ are phases arising from \FH{winding $n_j(\gamma)$ times around possible fluxes $\varphi_j$ \cite{Avron1994a}} (or a suitable generalization in higher dimensions),
\begin{equation}
    \FH{\chi_\gamma=\exp\left(i\sum_{j=1}^{2g}\varphi_jn_j(\gamma)\right),}
\end{equation}
and control time reversal invariance and thereby the symmetry class of the system: it is present when all $\chi_\gamma=1$ (orthogonal symmetry), and broken if any $\chi_\gamma\neq1$ (unitary), \FH{see also \cref{fig1}.}

\FH{In the remainder of this section, we will review precisely how the Plancherel measure enters \cref{eq:STF}, and show that it is playing the same role as the spectral density of JT gravity. Afterwards, we will take the limit of large configuration space dimension $f\to\infty$ and show that in this limit, $\Phi_f(p)$ tends exactly to the JT gravity result.}

\subsection{Plancherel measure}
In order to understand that the Plancherel measure in the STF has the same origin as the $\sinh$ spectral density in JT gravity, we need to recall some important steps in the derivation of the STF \cite{randol_selberg_1984}. The manifold $\mathcal{K}$ on which the dynamics plays out can be understood as a quotient manifold $\mathcal{K}=\mathds{H}^f/\Gamma$, where $\mathds{H}^f$ is the hyperbolic space in $f$ dimensions and $\Gamma$ some discrete hyperbolic subgroup of the group of orientation-preserving isometries on $\mathds{H}^f$. For $f=2,3$, $\Gamma$ would be a Fuchsian or Kleinian group respectively, and more generally, is isomorphic to the fundamental group $\pi_1(\mathcal{K})$, i.e. the set of closed geodesics on $\mathcal{K}$. Note that in the simplified notation of the rest of the text, $\Gamma$ refers only to the prime geodesics, or equivalently, the primitive elements of the fundamental group of $\mathcal{K}$, usually denoted in the literature by $\Gamma^*$. We consider the Laplacian $\Delta$ on $\mathds{H}^f$ and define the notion of point-pair invariants $k(x,y)=k(d(x,y))$, i.e. smooth, even functions of compact support, which depend only on the distance between two points $(x,y)\in\mathbb{H}^f\times\mathbb{H}^f$. Suppose now that we have some radial eigenfunction of the Laplacian, \emph{viz.} a function $\phi$ satisfying $\Delta\phi=\lambda\phi$ for some $\lambda$ and radial about a point $p$. Then for such a $\phi$, it is not difficult to show that
\begin{equation}
  \int_{\mathbb{H}^f}k(x,y)\phi(y)dy=h(\lambda)\phi(x), \label{eq:prePreTrace}
\end{equation}
with a function $h(\lambda)$ that does not depend on $\phi$. The left-hand side of \eqref{eq:prePreTrace} can be expressed as an integral over the fundamental domain $F$ of $\Gamma$, since $\Gamma F=\mathds{H}^f$:
\begin{equation}
  \int_{\mathbb H^f}k(x,y)\phi(y)dy=\int_F\left(\sum_{\gamma\in\Gamma}k(x,\gamma y)\right)\phi(y)dy, \label{eq:fundDomain}
\end{equation}
and combining Eqs. \eqref{eq:prePreTrace} and \eqref{eq:fundDomain} for a complete set of eigenfunctions $\phi_0,\phi_1,\phi_2,\ldots$, we recover Selberg's pretrace formula \footnote{\FH{The argument is perhaps a bit more subtle than a first glance would suggest. One needs to recognize that the LHS of \eqref{eq:preTrace} generates a Hilbert-Schmidt operator whose action is given by the RHS of \eqref{eq:fundDomain}. According to \eqref{eq:prePreTrace} then, the $\phi_n$ are eigenfunctions of this operator, and hence by Hilbert-Schmidt theory, the kernel of the operator admits an expansion as the RHS of \eqref{eq:preTrace}.}},
\begin{equation}
  \sum_\gamma k(x,\gamma y)=\sum_n h(\lambda_n)\phi_n(x)\overline{\phi}_n(y).\label{eq:preTrace}
\end{equation}
Taking the trace in \cref{eq:preTrace} gives the STF, and the term on the LHS coming from the conjugacy class of the identity $\gamma=\{\mathds{1}\}$ takes the form of the Weyl term
\begin{equation}
  \mathcal{V}\int_0^\infty h(\lambda)\Phi_f(\lambda)d\lambda,
\end{equation}
with the Plancherel measure, explicitly given by \cite{randol_selberg_1984}
\begin{equation}
  \Phi_f(p)=\frac{f}{(4\pi)^{f/2}\Gamma(\frac{f+2}{2})}\frac{\abs{\Gamma(ip+(f-1)/2}^2}{\abs{\Gamma(ip)}^2}, 
\label{eq:plancherel}
\end{equation}
coming in as the density of the radial eigenfunctions traced out on the RHS of \eqref{eq:preTrace},
\begin{equation}
  \int_{\mathds{H}^f}\phi_p(x)\overline{\phi}_{p'}(x)=\frac{1}{\Phi_f(p)}\delta(p-p'),\label{eq:selbergOrtho}
\end{equation}
where now $p$ is a variable parameterizing the continuous part of the spectrum of $\Delta$ and $\Delta\phi_p=\lambda(p)\phi_p$ \cite{Bytsenko1995}.

To see that the origin of the spectral density of JT gravity can be understood in the same way, we need to recall its formulation as a $\mathfrak{sl}(2,\mathds{R})$ $BF$ gauge theory. This theory is defined by the action
\begin{equation}
  S_{BF}=-i\int\tr(BF),
\end{equation}
where $B$ is a scalar field and $F$ the field strength computed from a gauge connection $A$. The integration is to be performed over the (as yet unspecified) gauge group manifold. This action can be straightforwardly mapped to the JT gravity bulk dilaton action, while the action of the boundary mode depends on the global structure of the gauge group, rather than just the gauge algebra. Taking the gauge group to be the universal cover $\widetilde{SL}(2,\mathds{R})$ of groups with the Lie algebra $\mathfrak{sl}(2,\mathds{R})$, the disk partition function can generically be expressed as an integral over the irreducible representations (irreps) of $\widetilde{SL}(2,\mathds{R})$ \cite{iliesiu_exact_2019},
\begin{equation}
  Z_{\text{disk}}\propto\int dR\Phi(R)e^{-\frac{\beta}{2C}\left[C_2(R)-\frac{1}{4}\right]},
\end{equation}
with the quadratic Casimir $C_2(R)$ and the Plancherel measure
\begin{equation}
  \Phi(R)dR=\frac{(2\pi)^{-2}s\sinh(2\pi s)}{\cosh(2\pi s)+\cos(2\pi\mu)}d\mu ds\label{eq:JTPlancherel}.
\end{equation}
Analytically continuing $\mu\to i\infty$ yields the $\sinh$ spectral density, and hence the disk partition function of JT gravity. The Plancherel measure comes in, once again, as the density of the basis functions of the ``irrep space'', namely the irreps $U_R$ themselves:
\begin{equation}
  (U_R,U_{R'})=\frac{1}{\Phi(R)}\delta(R,R'),\label{eq:groupOrtho}
\end{equation}
where $\delta(R,R')$ is the product of a Dirac (Kronecker) $\delta$ for every continuous (discrete) representation label in $R$. The correspondence between Eqs. \eqref{eq:selbergOrtho} and \eqref{eq:groupOrtho}, and hence of the Weyl term in the STF and the disk partition function of JT gravity, including the functional form of the Plancherel measure suggests a tantalizing unexplored connection between group quotients of high-dimensional hyperbolic spaces $\mathds{H}^f/\Gamma$ and groups like the gauge group of JT gravity, which can be understood either as a purely hyperbolic sector of a central extension of $\widetilde{SL}(2,\mathds{R})$ \cite{iliesiu_exact_2019}, or alternatively, a positive sub-semigroup of $SL(2,\mathds{R})$ \cite{blommaert_schwarzian_2018,blommaert_fine_2019}.
\subsection{The limit $f\to\infty$}\label{subsec:limit}
We now compute the limit of the Plancherel density as the dimension $f\to\infty$, which will allow us to show in \cref{sec:semiclass} that it agrees with the spectral density of JT gravity. To this end, it is helpful to distinguish between odd and even $f$,
\begin{widetext}
\begin{equation}
    \Phi_f(p)=\frac{f}{(4\pi)^{f/2}\Gamma(\frac{f+2}{2})}\frac{\abs{\Gamma(ip+(f-1)/2}^2}{\abs{\Gamma(ip)}^2}=
    \begin{dcases}
        \frac{p\tanh(\pi p)}{(2\pi)^{f/2}(f-2)!!}\prod_{k=0}^{\frac{f-4}{2}}\left(p^2+\left(k+\frac{1}{2}\right)^2\right) & f\text{~even}\\
        \frac{1}{2^{(f-1)/2}\pi^{(f+1)/2}(f-2)!!}\prod_{k=0}^{\frac{f-3}{2}}(p^2+k^2) & f\text{~odd}.
    \end{dcases}
\end{equation}
\end{widetext}
We start with the odd case. Separating the $k=0$ term and factoring out $k^2$ from the terms, we obtain
\begin{equation}\label{eq:phiOdd}
    \Phi_{f,\text{\,odd}}(p)=\frac{\frac{f-1}{2}!\left(\frac{f-3}{2}!\right)^2}{\pi^{(f+3)/2}(f-1)!}p\left(\pi p\prod_{k=1}^{\frac{f-3}{2}}\left(1+\frac{p^2}{k^2}\right)\right).
\end{equation}
Comparing to the infinite product representation 
\begin{equation}
    \sinh x = x\prod_{k=1}^{\infty}\left(1+\frac{x^2}{k^2\pi^2}\right),
\end{equation}
we identify
\begin{equation}
    \Phi_{f\to\infty,\text{\,odd}}(p)=\frac{p}{2\pi^2}\mathcal{N}_{\infty,\text{\,odd}}\sinh(\pi p),
\end{equation}
with the normalization constant $\mathcal{N}_{\infty,\text{\,odd}}=\lim_{f\to\infty}\mathcal{N}_{f,\text{\,odd}}$,
\begin{equation}\label{eq:normOdd}
    \mathcal{N}_{f,\text{\,odd}}=\frac{2\frac{f-1}{2}!\left(\frac{f-3}{2}!\right)^2}{\pi^{(f-1)/2}(f-1)!}.
\end{equation}
Notably, while the limiting expression grows exponentially with $p$, the Plancherel measure at every finite $f$ is simply a polynomial in $p$. 

In the even case, we can repeat essentially the same calculation, factoring out $(k+1/2)^2$, to find
\begin{equation}
    \Phi_{f,\text{\,even}}(p)=\frac{\Gamma\left(\frac{f-1}{2}\right)^2p\tanh(\pi p)}{\pi(2\pi)^{f/2}(f-2)!!}\prod_{k=0}^{\frac{f-4}{2}}\left(1+\frac{p^2}{\left(k+\frac{1}{2}\right)^2}\right),
\end{equation}
and comparing once more to the infinite product representation
\begin{equation}
\cosh x=\prod_{k=0}^\infty\left(1+\frac{x^2}{(k+1/2)^2\pi^2}\right),
\end{equation}
we find
\begin{equation}
    \Phi_{f\to\infty,\text{\,even}}(p)=\frac{p}{2\pi^2}\mathcal{N}_{\infty,\text{\,even}}\tanh(\pi p)\cosh(\pi p),
\end{equation}
with the normalization constant $\mathcal{N}_{\infty,\text{\,even}}=\lim_{f\to\infty}\mathcal{N}_{f,\text{\,even}}$,
\begin{equation}\label{eq:normEven}
    \mathcal{N}_{f,\text{\,even}}=\frac{\Gamma\left(\frac{f-1}{2}\right)^2}{(2\pi)^{\frac{f-2}{2}}(f-2)!!}.
\end{equation}
Obviously, the results for even and odd $f$ agree in the limit $f\to\infty$, but the approach to the limit is quite different, since the Plancherel density for even $f$ is not simply a polynomial. As a final, perhaps interesting observation, note that
\begin{equation}
    \frac{\mathcal{N}_{f,\text{\,even}}}{\mathcal{N}_{f,\text{\,odd}}}=\left(\frac{\pi}{2}\right)^{\frac{1}{2}\sin\left(\frac{f\pi}{2}\right)^2},
\end{equation}
i.e. the constants agree for even $f$ and differ by $\sqrt{\pi/2}$ for odd $f$.

\section{Semiclassical computation of correlation functions}\label{sec:semiclass}
\FH{Having constructed a class of systems to compare to JT gravity in the previous section, we now move on to computing correlation functions of the time evolution operator in such systems using semiclassical methods. As we will see, these computations will have some success in capturing the leading-topology JT one- and two-point functions at least in some regimes, but fail in others.}
\subsection{Weyl term and Periodic Orbit Sum}
In a usual semiclassical treatment, the Weyl term would be estimated by a Thomas-Fermi approximation $\rho_{f}^{\text{TF}}(p^2)$, i.e. an integral over an energy shell in phase space \cite{Balian1971, brack2003semiclassical}. \FH{For the class of systems introduced in \cref{sec:model},} this agrees with the exact density $\Phi_f(p)$ at large energies,
\begin{equation}
  {\cal V}\Phi_{f}(p) =\rho_{f}^{\text{TF}}(p^{2})\left(1+{\cal O}(1/p)\right) \, .
\end{equation}
The Plancherel measure $\Phi_f(p)$ can then be understood to contain \textit{all quantum corrections} to $\rho_{f}^{\text{TF}}(p^2)$ \cite{Balazs1986}. 

\FH{Taking the limit $f\to\infty$ as discussed in \cref{subsec:limit}, and}
absorbing the \FH{normalization constant \eqref{eq:normOdd}, resp. \eqref{eq:normEven},} into the appropriately chosen volume ${\cal V}$, the spectral density appearing in \cref{eq:STF} finally reads 
\begin{equation}
    \FH{\mathcal{V}\Phi_\infty(p)=\sinh{(\pi p)}.}
\end{equation} 
Understanding $f$ as the number of configuration space degrees of freedom, this limit is tantalizingly reminiscent of the double scaled limit in which the matrix model of \cite{Saad2019} agrees with JT gravity.

To proceed further, we need to specify the spectral function $h(p)$. In the JT gravity matrix model \cite{Saad2019}, the operator inserted on the matrix side to compute the JT gravity path integral with standard boundary conditions is the trace of the heat kernel. In our approach, the appropriate choice turns out to be essentially the same, namely $h(p)=e^{-i\tau p^2/4}$. When summed over the spectrum of the Hamiltonian, this choice yields the trace of the (analytically continued) heat kernel up to an interesting scaling of the complex time $\tau=t-i\beta, \beta > 0$ by a factor of four~\footnote{This choice of spectral function, together with our earlier choice of units, allows us to use standard results for the heat kernel ${\rm Tr}~{\rm e}^{-\beta H}$ for lengths measured in units of $L$, since these depend only on $l^2/\tau$.}. Plugging this into \cref{eq:STF} yields for the Weyl term
\begin{equation}
  \begin{aligned}\mathcal{V}&\int_0^\infty \frac{p\,dp}{2\pi^2}e^{-i\tau p^2/4}\sinh{\pi p}\\&=\mathcal{V}\int_0^\infty e^{-i\tau E}\frac{1}{4\pi^2}\sinh(2\pi\sqrt{E})dE \, ,
  \end{aligned}
\end{equation}
\FH{which} is exactly the JT gravity disk partition function including, crucially, the $\sinh$-type spectral density \eqref{eq:sinh}. It is interesting to note that, up to factors absorbed into the choice of ${\cal V}$, $\Phi_{f=3}(p)\propto p^2$ exactly reproduces the spectral density of the Airy model \cite{Kontsevich1992}. The Weyl term of the STF therefore interpolates between the Airy and JT spectral density for $f=3$ and $\infty$, in a way akin, but not identical to the $(2,q)$ minimal string for $q=1,\infty$ \cite{Saad2019,Mertens2021}. \FH{In particular, \eqref{eq:phiOdd} is a polynomial of degree $f-2$ in $p$ for any odd $f$, while the minimal string spectral density is a polynomial of degree $q$ (indeed the difference of two Chebyshev polynomials \cite{Gregori2021}). It would be tempting to identify $q=f-2$ and conclude that our model possesses a dual description as a minimal string theory in any (odd) dimension, but the spectral densities only agree for $f-2=q=1$ and for $f\to\infty,q\to\infty$.}

\FH{We see then that decomposing Eq. \eqref{eq:STF} as}
\begin{equation}
Z(\tau)=\tr(e^{-i\tau \hat{H}/4})=Z_{\text{Weyl}}(\tau)+Z_{\text{PO}}(\tau) \, , 
\label{eq:Z}
\end{equation}
\FH{the Weyl term is simply the gravitational correlator $Z_{0,1}$, while we have yet to do anything}
with the periodic orbit contribution 
\begin{equation}
\label{eq:Ztilde} 
  Z_{\text{PO}}
  (\tau)=\sum_{\gamma\in\Gamma}\sum_{m=1}^{\infty}\frac{\chi_{\gamma}^{m}l_\gamma}{S_f(m,l_\gamma)}\frac{{e}^{-\frac{m^{2}l_{\gamma}^{2}}{i\tau}}}{\sqrt{\pi i\tau}}.
\end{equation}

Note the remarkable appearance of $Z^{\rm t}_{\rm Sch}(i\tau,l_{\gamma})$, Eq.~(\ref{eq:trumpet}), with its exponent now admitting an interpretation as the classical action of the periodic orbit $\gamma$. \FH{We will show in the next subsection that the standard semiclassical computation of the connected two-point correlation function of \eqref{eq:Z}, or equivalently of \eqref{eq:Ztilde}, gives the double trumpet correlation function of JT gravity in the ramp regime. To go beyond this regime, we will introduce a modified way of computing correlation functions in \cref{sec:avg}.}

\subsection{Correlation functions}
Assuming $\abs{t} \gg \beta$,
\FH{the} exponents \FH{in \eqref{eq:Ztilde}} give rise to highly oscillatory contributions to $Z_{\rm PO}(\tau)$. 
Usually, correlation functions involving periodic orbit sums are semiclassically evaluated through a \textit{smoothing} procedure that acts on \textit{oscillatory} expressions, such as \cref{eq:Ztilde}. Thereby, quantum interference between amplitudes associated with so-called correlated periodic orbits, surviving this smoothing, is highlighted \cite{Sieber2003,richter_semiclassical_2022,Muller2005}. Note that this average, hereinafter denoted as $\ev{\cdot}_{\rm sc}$, operates on the level of a \emph{single} system, without the necessity to introduce disorder averages 
or some other type of ensemble of systems.

Accordingly, we define semiclassical connected correlators of partition functions $Z(\tau)$, \cref{eq:Z}, through the smoothing of (products of) their periodic-orbit terms (\ref{eq:Ztilde})
(for $\abs{t} \gg \beta$) and identify contributions to the single and double sums that do not vanish under $\ev{\cdot}_{\rm sc}$. For the 1-point function, this simply means 
\begin{equation}
\ev{Z(\tau)}_{\rm sc}=Z_{\rm Weyl}(\tau)=Z_{0,1}(i\tau) \, ,
\end{equation}
since $Z_{\rm Weyl}$ is by definition smooth and we have shown $Z_{\rm Weyl}=Z_{\rm Sch}^{\rm d}$. However, the semiclassical (connected) 2-point function, 
\begin{align}
  &{\cal Z}_{\rm sc}^{\rm c}(\tau_{1},\tau_{2})=\langle Z_{\text{PO}}(\tau_{1})Z_{\text{PO}}(\tau_{2})\rangle_{\rm sc}\\
  &\FH{=\sum_{\gamma,\gamma^\prime}\sum_{m,m^\prime}\frac{\chi_{\gamma}^{m}l_\gamma}{S_f(m,l_\gamma)}\frac{\chi_{\gamma^\prime}^{m^\prime}l_{\gamma^\prime}}{S_f(m^\prime,l_{\gamma^\prime})}
  \frac{{e}^{-\frac{m^{2}l_{\gamma}^{2}}{i\tau_1}}}{\sqrt{\pi i\tau_1}}\frac{{e}^{-\frac{(m^\prime)^{2}l_{\gamma^\prime}^{2}}{i\tau_2}}}{\sqrt{\pi i\tau_2}}}
\end{align}
critically depends on the relative sign of the time variables $t_{1},t_{2}$, which controls the existence of slow oscillatory contributions arising from quantum interference. Inspection of the products involved then shows 
\begin{equation}
  {\cal Z}_{\rm sc}^{\rm c}(\tau_{1},\tau_{2})= 
  \begin{cases}
  0 & {\rm for \ } t_{1}t_{2} > 0, \\
  {\cal Z}_{\rm sc}^{\rm corr}(\tau_{1},\tau_{2}) & {\rm for \ } t_{1} \simeq -t_{2},
  \end{cases}
  \label{eq:cases}
\end{equation}
where ${\cal Z}_{\rm sc}^{\rm corr}$ comprises pairs of \textit{correlated} orbits with systematically small action differences \cite{Berry1985a, Sieber2003}. 
A well established result of the semiclassical analysis, Berry's diagonal approximation \cite{Berry1985a}, is that for times shorter than a scale set by $\rho_{0}$, 
the double sum is dominated by its diagonal part, i.e., ${\cal Z}_{\rm sc}^{\rm corr}\simeq {\cal Z}_{\rm sc}^{\rm dg}$, where
\begin{equation}
\label{eq:diag}
 {\cal Z}_{\rm sc}^{\rm dg}(\tau_{1},\tau_{2})= \sum_{\gamma}\sum_{m=1}^{\infty}\frac{\kappa l_{\gamma}^{2}}{S_{f}^{2}(m,l_\gamma)}\frac{{e}^{-\frac{m^{2}l_{\gamma}^{2}}{i\tau_{1}}}}{\sqrt{\pi i\tau_{1}}} \frac{{e}^{-\frac{m^{2}l_{\gamma}^{2}}{i\tau_{2}}}}{\sqrt{\pi i\tau_{2}}} \, .
\end{equation}
\FH{Note the appearance of only one sum over periodic orbits in Eq. \eqref{eq:diag}.}
This \FH{result} is corrected by
 off-diagonal terms \cite{Sieber2003,Muller2004} representing quantum interference effects. The 
presence (absence) of time-reversal-symmetry, encoded in $\kappa=2(1)$ \cite{Berry1986}, is 
responsible for the first (weak localization) correction.

As long as $i\tau_{1} \simeq (i\tau_{2})^*$, the function ${\cal Z}_{\rm sc}^{\rm dg}(\tau_{1},\tau_{2})$ is by definition smooth. Thus, the sum \FH{over periodic orbits} can be written as an integral by introducing the weak limit $\bar{\eta}(l)$ of the density
\begin{equation}
  \eta(l):=\sum_{\gamma}\delta(l-l_{\gamma}) \stackrel{{\rm weak}}{=}\bar{\eta}(l)=
  \frac{2\sinh{\left[(f-1)l\right]}}{l} 
  \label{eq:density}
\end{equation}
 of closed geodesic lengths.
The last equality \FH{expresses} the exponential proliferation of closed geodesics \cite{BOGOMOLNY1997,STFbook}, valid for $l>l_{0}$ where $l_{0}$ is set by the shortest one. Using the explicit form $S_{f}(m,l)=(2\sinh^{m}(l/2))^{(f-1)}$\cite{Bytsenko1995,randol_selberg_1984} for the stabilities in \cref{eq:Ztilde}, and known results in hyperbolic geometry \cite{Sytole2011} implying $l_{0} \gg 1/{\cal V}^{1/f}$, within the regime of integration, the decaying squared stability factors exactly cancel the exponential proliferation,
\begin{equation}\label{eq:expCancellation}
    \frac{\bar{\eta}(l) l}{S_f^2(1,l)} \to 1 {\rm \ \ for \ \ } l > l_{0},
\end{equation}
but only for $m=1$. Terms with $m \ge 2$ are also suppressed since the number of primitive geodesics grows exponentially faster than the number of repetitions \cite{BOGOMOLNY1997}. 

We now quantify the range of $\tau_{1,2}=t_{1,2}-i\beta_{1,2}$ where \cref{eq:diag} is valid\FH{. To this end, define} $t_{1}=t-\epsilon/2, t_{2}=-t-\epsilon/2$. Using semiclassical techniques \cite{Berry1985a}, we expand the actions in Eq.~(\ref{eq:diag}) to first order in $\epsilon/t,\beta_{1,2}/t$, evaluate all prefactors at zeroth order and extend the integration range, to get
\begin{equation}
\label{eq:Zdiag}
\begin{aligned}
  {\cal Z}_{\rm sc}^{\rm dg}\left(t-\frac{\epsilon}{2}-i\beta_{1}, -t-\frac{\epsilon}{2}-i\beta_{2}\right)&= \\ \kappa\int_{0}^{\infty}\frac{ldl {\rm e}^{-\frac{{l}^{2}}{2t^{2}}(\bar{\beta}+i\epsilon)}}{2\pi \abs{t}}+{\cal O}\left(\frac{\beta_{1,2}}{t},\frac{\epsilon}{t}\right)&\simeq\frac{\kappa \abs{t}}{2\pi}\frac{1}{\bar{\beta}+i\epsilon}
\end{aligned}
\end{equation}
with $\bar{\beta}=\beta_{1}+\beta_{2}$. This result contains in particular the ramp \FH{\eqref{eq:ramp}} valid for $\abs{t} \gg \beta$ (and reported for $f=2,\kappa=1$ in \cite{Garcia-Garcia2019}). There can be symmetry class-dependent universal corrections to \cref{eq:Zdiag}, and our system (indeed any chaotic quantum system) can capture these as well, as shown by \cite{Sieber2003,Muller2005}. They do not appear in \cref{eq:Zdiag} however, as they correspond to higher topology, while we are only concerned with the genus 0 contribution to the 2-point function here.

To summarize, the diagonal approximation yields the gravitational result for $Z_{0,2}$ for $t_{1}\simeq -t_{2}$ and $|t_{1,2}| \gg \beta_{1,2}$ while it fails in any other regime, see \cref{eq:cases}. \FH{As we shall see in the next section, replacing the usual semiclassical smoothing by a more suitable coarse graining in the phase space allows to go beyond this regime and recover the full double trumpet \eqref{eq:Z02} for any values of $\tau_1,\tau_2$.} \footnote{Interestingly, computations in de Sitter JT gravity suggest that $t_{1}\simeq -t_{2}$ is the only classical configuration of the theory (global dS$_2$), and therefore is enhanced \cite{Cotler2020}. However, as stressed in \cite{Saad2019}, this is not the case for AdS JT of interest here.}.

\begin{figure}[ttt]
\centering
\includegraphics[width=0.66\linewidth]{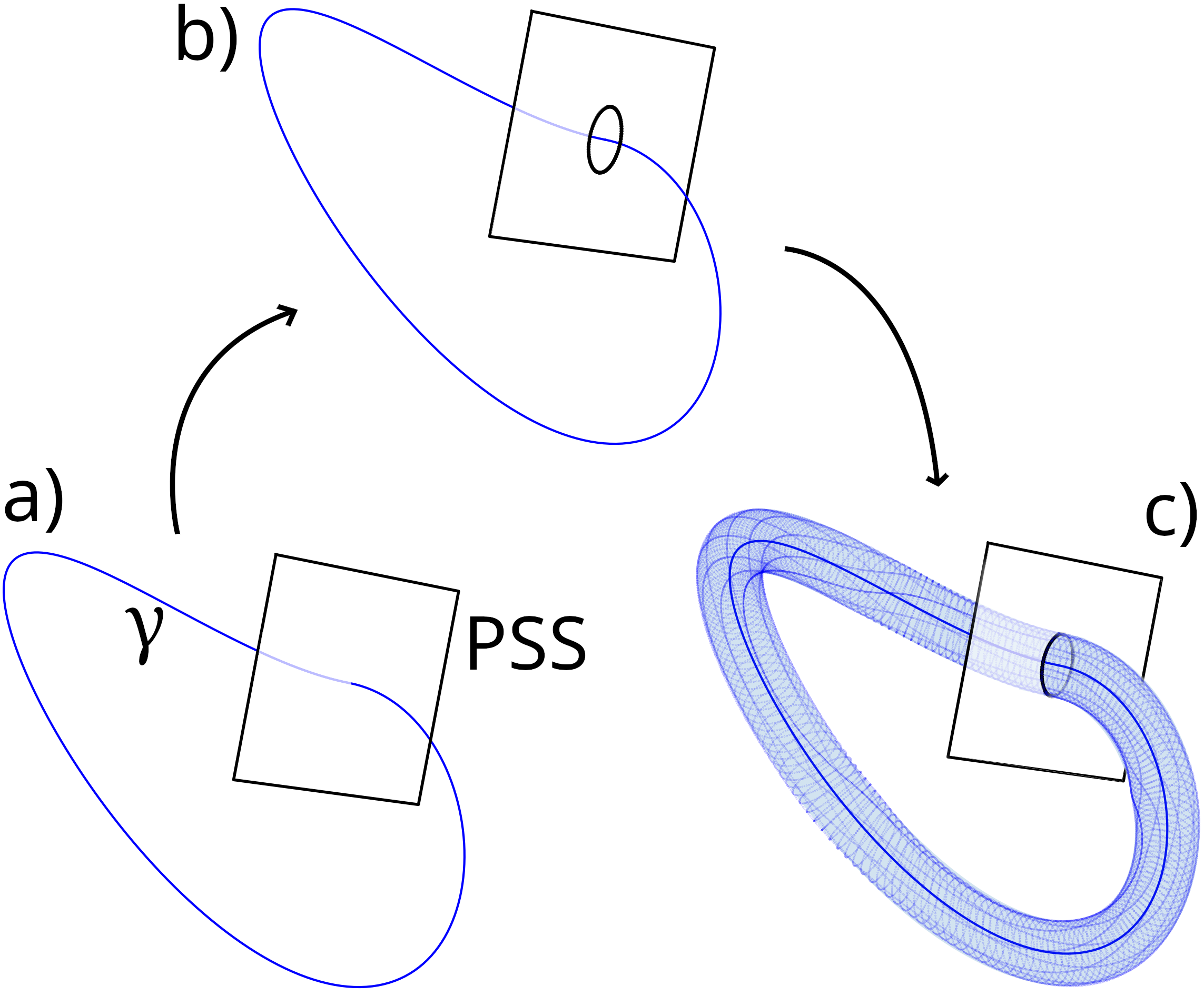}
\caption{a) Every periodic orbit $\gamma$ in phase space is determined by the points where it pierces the Poincare Surface of Section (PSS). b) When a finite resolution on the PSS is introduced, any property of the periodic orbits, like the joint distribution of their lengths, (c) cannot be determined with infinite precision.}
  \label{fig2}
\end{figure}

\section{Length average}\label{sec:avg}
 Aiming to reconcile the semiclassical and gravitational results for $Z_{0,2}$ beyond the universal regime, we reconsider the notion of smoothing average again, now from the general perspective of the loss of information that takes place when neglecting highly oscillatory contributions in sums over amplitudes. To go beyond this picture, but still such that we define the correlation functions for a single individual system, we follow standard methods of statistical physics \cite{grandy1988} and introduce a \textit{coarse graining} description. This coarse graining is implemented by introducing a finite action scale, blurring structures in phase space, with the consequent loss of information. The corresponding probabilistic description is then a consequence of our ignorance, rather than the existence of a physical ensemble such as in disordered systems. While our ideas get support from the recent proposal in \cite{Sonner2024} based on stochastic methods, it would be desirable to identify the dynamical origin of the information loss in the spirit of \cite{Post2022,Altland2022} where JT correlators are derived within the ``universe'' (Kodaira-Spencer) field theory approach as expectation values of operators insensitive to certain microscopic degrees of freedom that are precisely identified by the theory. 

 Considering for simplicity the orthogonal ($\kappa=2$) case of $\chi_{\gamma}=1$ in Eq.~(\ref{eq:STF}), we implement the coarse graining as follows. To specify a periodic orbit in phase space, it is enough to know the points where it pierces 
any manifold transversal to the classical flow \cite{AOdA}, called Poincar\'{e} surface of section. As illustrated in \cref{fig2}, a finite resolution on this manifold introduces a degree of ignorance about the orbits. In particular, the exact microscopic (mc) distribution $P_{\bf{l}(\Gamma)}^{\rm mc}$ of lengths ${\bf l}=(l^{(1)},l^{(2)},\ldots)$,
\begin{equation}
  P_{{\bf l}(\Gamma)}^{\rm mc}({\bf l})=\prod_{\gamma}\delta(l^{(\gamma)}-l_{\gamma}), {\rm \ with \ }{\bf l}(\Gamma)=(l_{1},l_{2} \ldots) \, , 
\end{equation}
will no longer be sharp. Its coarse-grained (cg) version $P^{\rm (cg)}$ provides instead a statistical description of the length set, including the systematic correlations rigorously shown to exist \cite{Huynh2015a,Huynh2015b} and to be responsible for the emergence of universal RMT spectral fluctuations \cite{Muller2004}. As any object built upon the length spectrum, the length density $\eta(l)$ is no longer sharp in the $l_{\gamma}$'s but described by its statistical fluctuations under the coarse-grained distribution. In particular, its first two moments read
\begin{equation}
\label{eq:etacorr}
  \begin{aligned}
    \ev{\eta(l)}&=\sum_{\gamma}\int d{\bf l}P^{\rm cg}({\bf l})\delta(l-l_{\gamma})\, , \\ 
\ev{\eta(l)\eta(l')}&=\sum_{\gamma,\delta}\int d{\bf l}P^{\rm cg}({\bf l})\delta(l-l^{(\gamma)}) \delta( l'-l^{(\delta)}) \, .
  \end{aligned}
\end{equation}
By means of the corresponding (c)onnected correlator
\begin{equation}
\label{eq:conn}
\ev{\eta(l)\eta(l')}^{\rm c}=\ev{\eta(l)\eta(l')}-\ev{\eta(l)}\ev{\eta(l')}
\end{equation}
the connected 2-point correlation function of periodic orbit sums (\ref{eq:Ztilde}) 
 is written, in view of Eq.~(\ref{eq:trumpet}), as 
\begin{eqnarray}
\label{eq:Zconn}
  \langle Z_{\text{PO}}(\tau_{1})Z_{\text{PO}}(\tau_{2})\rangle^{\rm c} &=&\int_{0}^{\infty}\frac{ldl}{S_{f}(1,l)}\frac{l'dl'}{S_{f}(1,l')}\ev{\eta(l)\eta(l')}^{\rm c} \nonumber \\ &&\times Z^{\rm t}_{\rm Sch}(i\tau_{1},l) Z^{\rm t}_{\rm Sch}(i\tau_{2},l') \, .
\end{eqnarray}
Here, consistently with the derivation of Eq.~(\ref{eq:Zdiag}), terms with $m\ge 2$ have been neglected. Note that in the equations above, the length average $\ev{\cdot}$ is employed \emph{instead of} $\ev{\cdot}_{\text{sc}}$, not \emph{on top of it}.

After having specified the way statistical properties of the set of lengths determine the semiclassical correlation functions, we ask for the properties of $P^{\rm cg}({\bf l})$. Following a standard approach of periodic orbit theory, we start by considering the lengths as uncorrelated random variables \cite{BogKeat1996}. The self-correlations, however, have a distinct role when (as in our case) the probability distribution is defined over a discrete space. For uncorrelated variables, the definitions \eqref{eq:etacorr}, where the diagonal term $\gamma=\delta$ produces a singular contribution, imply the existence of contact terms in correlators of the form \eqref{eq:conn} \cite{mehta2004}. Invoking the fundamental property of chaotic systems that the set of periodic orbits is \textit{discrete} \cite{AOdA} and the straightforward condition $\ev{\eta(l)}=\bar{\eta}(l)$, we get
\begin{equation}
\ev{\eta(l)\eta(l')}_{\rm un}^{\rm c}=\kappa \bar{\eta}(l) \delta(l-l') \, 
\end{equation}
for uncorrelated lengths,
 where $\kappa$ is reintroduced following the analysis of \cite{Berry1986}. Substitution in Eq.~(\ref{eq:Zconn}) yields
 \begin{equation}
   \begin{aligned}
     &\langle Z_{\text{PO}}(\tau_{1})Z_{\text{PO}}(\tau_{2})\rangle_{\rm un}^{\rm c} \\ &= \kappa \int_{0}^{\infty}\frac{\bar{\eta}(l)l^{2}dl}{S_{f}^{2}(1,l)}Z^{\rm t}_{\rm Sch}(i\tau_{1},l) Z^{\rm t}_{\rm Sch}(i\tau_{2},l) \, . 
   \end{aligned}\label{eq:twopoint}
 \end{equation}
Finally, using \FH{Eq. \eqref{eq:expCancellation} in a similar manner as in the derivation of} Eq.~(\ref{eq:Zdiag}), we \FH{reproduce the full double trumpet partition function} \eqref{eq:Z02},
\begin{equation}
\label{eq:Z02sc}
  \langle Z_{\text{PO}}(\tau_{1})Z_{\text{PO}}(\tau_{2})\rangle_{\rm un}^{\rm c}=\kappa Z_{0,2}(i\tau_{1},i\tau_{2}) \, . 
\end{equation}
This result holds for \textit{all complex values of $\tau_{1,2}$}, in particular for the universal regime $t_{1}\simeq -t_{2}, \abs{t_{1,2}} \gg \beta_{1,2}$, where it coincides with the diagonal approximation, \cref{eq:Zdiag}.
Moreover, our approach goes beyond the universal result and identifying the periodic orbit length $l$ with the length of the internal gluing geodesics $b$ appearing in the computation of JT gravity partition functions, Eq.~(\ref{eq:Z02}), Eq.~(\ref{eq:Z02sc}) gives the exact result for the JT ``double trumpet'' or wormhole partition function. This agreement extends well beyond the ramp, Eq.~(\ref{eq:Zdiag}) as it holds at the more fundamental level of its full integral form in Eq.~(\ref{eq:Z02}). In this way the Weil-Petersson measure $ldl$, the Schwarzian trumpets of \cite{Saad2019} and the $\kappa$ index reflecting the orientability of the manifolds in the gravitational genus expansion, all admit a periodic orbit interpretation.

\section{Intermission: Our notion of duality}
\FH{At this point, readers more familiar with holography in higher dimensions may ask the question of whether we can prove an exact duality between our model and JT gravity, in the sense of identifying the two Hilbert spaces, as well as the spectra of the respective Hamiltonians. The short answer is no, but it is our understanding that this is not the usual notion of duality that is employed in most of the literature studying JT gravity. In particular, JT gravity is dual to the random matrix ensemble of \cite{Saad2019} and as such, does not have a fixed, concrete energy spectrum, but rather a spectrum that is an average over said ensemble, with the smoothed spectral density \eqref{eq:sinh}. In order to obtain a concrete spectrum, one would have to fix a specific member of the JT gravity ensemble, or equivalently, a specific $\alpha$ state in the baby universe Hilbert space of JT gravity \cite{Marolf_2021}. Such a fixing procedure would require the insertion a large number of fixed-energy branes in the path integral \cite{Blommaert2019,Blommaert_2022}, or equivalently, $\delta$ functions fixing individual eigenvalues in the matrix integral, which leads to a highly nontrivial gravitational interpretation for a large number of fixed eigenvalues \cite{Blommaert2021}. Given a spectrum fixed in this way, one could then try to find a concrete high-dimensional manifold whose Laplacian spectrum matches the one of our would-be gravitational theory, i.e. the specific member of the JT ensemble. In practice, this problem seems to be out of reach however.}

\FH{The Hilbert space of JT gravity on the other hand can be defined even without specifying a member of the ensemble, as e.g. in \cite{Iliesiu_Levine_Lin_Maxfield_Mezei_2024}. This Hilbert space, parametrized by the lengths of certain geodesics in the spacetimes appearing in the path integral, and containing a large number of null states, seems at first sight quite different from the natural Hilbert space of our model, namely $L^2(\mathcal{K})$, where $\mathcal{K}$ is our manifold. However, given that there are no natural candidates for null states in our model, it seems reasonable to assume that what we would see is the Hilbert space with null states already quotiented out, as argued in \cite{Blommaert_2022}. An interesting question to ask then is whether, similar to \cite{Blommaert_2022} and expected by \cite{Saad2021a}, there is some way to physically interpret the null states of \cite{Iliesiu_Levine_Lin_Maxfield_Mezei_2024} regardless, perhaps as different manifolds with the same spectrum. Given the close relation between the spectrum of the Laplacian and the geodesic length spectrum on hyperbolic manifolds, this is an intriguing possibility. In three dimensions for instance, it has been shown that there are infinitely many manifolds with a given initial geodesic length spectrum \cite{millichap_mutations_2017}.}

\FH{It appears very difficult then to prove a duality of any one system to JT gravity in the more strict notion common in higher-dimensional holography, but the advantage of our approach is that we nevertheless provide a mechanism to study holography in the more relaxed sense of the JT/RMT duality, i.e. at the level of correlation functions, \emph{while still only working with a single system}, rather than an ensemble.}

\section{Outlook: Genus expansions}\label{sec:genus}
\FH{The semiclassical theory of spectral correlations over small energy windows $[\bar{E}-e/2,\bar{E}+e/2]$, including effects beyond the diagonal (leading order) term, has been successfully formulated as a perturbative expansion in a series or remarkable works summarized in \cite{Muller2005},
 see also \cite{richter_semiclassical_2022} for a recent review from the many-body perspective.
After including non-perturbative effects, as done in \cite{Heusler2006}, this program correctly predicts the emergence of universal spectral fluctuations around the fixed scale $\bar{E}$ in quantum systems with chaotic classical limit. After the universal microcanonical results are derived from periodic orbit theory, a suitable Laplace transform from the energy $\bar{E}$ to the canonical domain, i.e. to the dependence on the inverse temperature $\beta$, that requires as input the system-dependent mean level density $\rho_{0}(\bar{E})$, gives in turn the desired canonical correlation functions, as shown for the unitary symmetry class in \cite{saad_convergent_2022} and the orthogonal in \cite{weber_unorientable_2024}. We therefore conclude that in the \textit{universal regime} the semiclassical approach, for the model we propose, gives results that are identical to the \textit{full} genus expansion of the JT correlation functions. For JT gravity and the Airy model, the contributions in the universal regime to said canonical correlation functions have been worked out in \cite{saad_convergent_2022, blommaert_integrable_2022,anegawa_late_2023, Weber_2023, Tall2024}.}

\FH{In this sense, the semiclassical approach is fully consistent with the gravitational correlators to all orders in the genus expansion as long as the universal regime is concerned. However, this is in fact a straightforward consequence of the universal spectral fluctuations of \textit{all} systems with chaotic classical limit. This agreement therefore does not shed any light on the much more demanding goal of finding a single chaotic system whose correlation functions agree with JT gravity beyond the universal regime.}

\FH{A possible path towards this fundamental question is to study the so-called encounter mechanism invoked in the semiclassical calculation of universal spectral correlations in the classic works \cite{Sieber2003,Muller2005} to see to what extent it can also be used to address the non-universal regime. This idea is the subject of present study \cite{futurework}.}

\section{Discussion}
To conclude, based on the Selberg trace formula we studied spectral correlations of quantized chaotic motion in high-dimensional manifolds of constant negative curvature and \FH{took initial steps towards establishing a duality} to JT gravity by deriving exact results for the leading-order terms of two JT correlation functions. First, we show the strict equivalence of the two theories at the level of the \FH{leading-topology} 1-point correlator where the Weyl term representing the smooth spectral density of the hyperbolic dynamical system exactly coincides with the gravitational result, as given by the Schwarzian action on a disk topology. Second, while under the usual semiclassical smoothing, the 2-point correlator of the hyperbolic system reproduces the JT result for $Z_{0,2}(i\tau_{1},i\tau_{2})$ only in the random matrix theory regime $\Im \tau_{1}\sim \Im \tau_{2} \gg \Re \tau_{1,2}$ (the so-called ramp contribution), the combination of phase space coarse graining and discreteness of the underlying periodic orbit set extends this result to the full wormhole geometry and all complex times. In this way, our approach provides a periodic orbit interpretation of distinct generic features on the gravitational side, notably at the level of an individual quantum system, as opposed to the ensemble paradigm (e.g. \cite{Marolf2020,marolf2024natureensemblesgravitationalpath} and references therein) in low-dimensional quantum gravity, with the large dimensionality of the hyperbolic system hinting towards chaotic many-body dynamics. 

In order to \FH{further strengthen our approach,} it is desirable to show whether the geodesic motion in high dimensional manifolds can be described by the Schwarzian action or to identify the telltale $SL(2,\mathds{R})$ symmetry, both characteristic properties of JT gravity also present in, for example, the SYK model. This, and the possible existence of a genus-like expansion on the semiclassical side emerging from the interplay between the encounter mechanism \cite{Sieber2003, Muller2004,
saad_convergent_2022} and the finite resolution introduced here, are subject of present scrutiny.

{\em Acknowledgments --} We thank A. Altland, A.~M. Garc\'{\i}a-Garc\'{\i}a and T. Mertens for useful conversations, and acknowledge financial support from the Deutsche Forschungsgemeinschaft (German Research Foundation) through project Ri681/15-1 (project number 456449460) within the Reinhart-Koselleck Program. We also thank F.~Schöppl for help in preparing the manuscript.

\bibliography{library_short.bib}
\end{document}